# WEAK-STRONG SIMULATION OF BEAM-BEAM EFFECTS IN SUPER PROTON-PROTON COLLIDER

Lijiao Wang†, [1], Jingyu Tang, IHEP, Beijing, China
Tanaji Sen, FNAL*, Batavia, IL 60510, USA
[1]visiting student at NIU, DeKalb, IL 60115, USA

*Abstract*

A Super Proton-Proton Collider (SPPC) that aims to explore new physics beyond the standard model will be built in China. Here we focus on the impact of beam-beam interactions in the SPPC. Simulations show that with the current optics and nominal tunes, the dynamic aperture (DA) with all the beam-beam interactions is less than 6σ, the dominant cause being the long-range interactions. First, we show the results of a tune scan done to maximize the DA. Next, we discuss the compensation of the long-range interactions by increasing the crossing angle and also by using current carrying wires.

## INTRODUCTION

The super proton proton collider (SPPC) is a 100km circular accelerator that aims to reach a luminosity of $1.01^{35}$cm$^{-2}$s$^{-1}$ [1]. Two proton beams with a design energy of 37.5Tev travel in opposite directions in separate beam pipes, then collide in 2 interaction regions (IR), one where the beams collide in the horizontal plane, and at the other in the vertical plane. A major limit to the luminosity arises from the beam beam effects, the nominal beam beam parameter for each IP is 0.0075. In addition to the head-on interaction at each IP, there are 82 long-range interactions in each IR [2]. Table 1 lists the main SPPC nominal parameters.

Table 1: SPPC nominal parameters

| parameter | value |
|---|---|
| Beam energy at collisions [TeV] | 37.5 |
| Number of bunches | 10080 |
| β∗ [m] | 0.75 |
| Crossing angle [μrad] | 110 |
| Intensity [$10^{11}$ p/bunch] | 1.5 |
| Norm. trans. emittance [μm] | 2.4 |
| Peak luminosity [$10^{35}$ cm$^{-2}$·s$^{-1}$] | 1.01 |

## TUNE FOOTPRINTS AND DYNAMIC APERTURE

In the nominal lattice design, the tune is (120.31, 117.32), the fractional parts being the same as in the LHC. At present, the only nonlinearities in the machine lattice are from the chromaticity correcting sextupoles. Figure 1 shows the beta functions in one IR. Figure 2 shows the plot of the LR separations at all the beam-beam interactions in each IR. The minimum separation is 9~10 σ, which occurs at 20 parasitic interactions in each IR. The weak-strong simulations reported here are done using the code BBSIM [3]. Figure 3 shows the tune footprint from the head-on and long-range interactions.

Dynamic aperture (DA) calculations are done in six dimensional phase space with particles tracked for a million turns. Tracking with only the sextupoles and the head-on interactions show that the DA is the same as the physical aperture of 23.6 σ. Adding the long-range interactions reduces the DA to 5.5σ, which is unacceptably low. Since this model does not yet include all the magnetic nonlinearities or errors, we have to find ways to increase the DA.

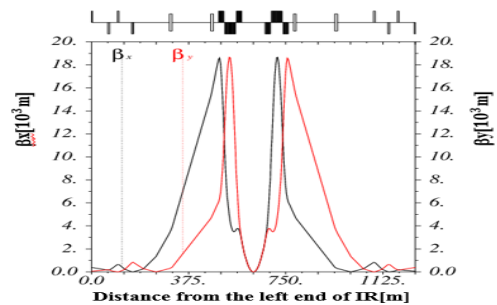

Figure 1: The beta functions in an IR.

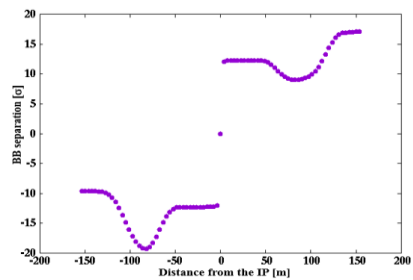

Figure 2: The BB separation of all LR interactions are normalised by its horizontal beam size in IR.

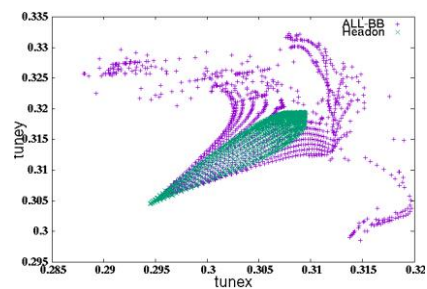

Figure 3: Green points is the tune footprint of head-on interactions, violet points is the tune footprint with all of interactions.

## TUNE SCAN STUDIES

We scanned the tunes in the range (0.10 ≤. $\nu x$ ≤ 0.46), ($\nu y = \nu x \pm 0.01$ or 0.02). Figure 4 shows the DA for different tunes, both without and with the nominal full crossing angle of 110 μrad. We can see 3rd, 4th, 5th order sum resonances are dangerous even without the crossing angle while with the crossing angle, 9th and 10th order sum resonances are also excited. In addition, with non-zero momentum deviation, synchro-betatron resonances will be excited by the presence of the crossing angle, which can reduce the DA further. Figure 5 shows that with increasing momentum deviation, there are more synchro-betatron spectrum lines excited and their amplitude become larger as well.

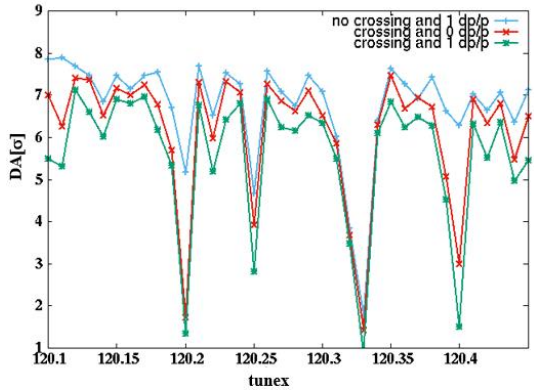

Figure 4: Average DA with different $\nu x$. $\nu y = \nu x \pm 0.01$; red; with crossing angle and no $\Delta p/p$ (momentum deviation); green: with crossing angle and $\Delta p/p = \sigma p$ (rms momentum spread); blue: without crossing angle and $\Delta p/p = \sigma p$

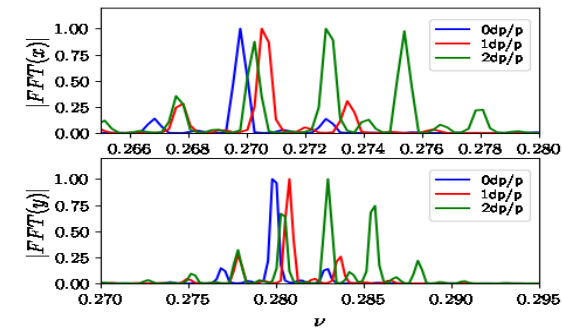

Figure 5: Horizontal and vertical spectrum for a particle with initial amplitude 1 $\sigma x$, 1 $\sigma y$ and 1 $\sigma z$ with different momentum deviations.

The tune scan reveals there are 4 tunes where the DA increases by 1.5σ to 7σ and 4 others where the increase in DA is ~1σ. The best tunes are (0.12, 0.13), (0.17, 0.19), (0.24, 0.26) and (0.27, 0.26). Figure 6 shows us the Frequency Map Analysis (FMA) [4] plot in amplitude space for the nominal tune and a better tune of (0.27, 0.26). We can observe that the tune variation is smaller at the better tune.

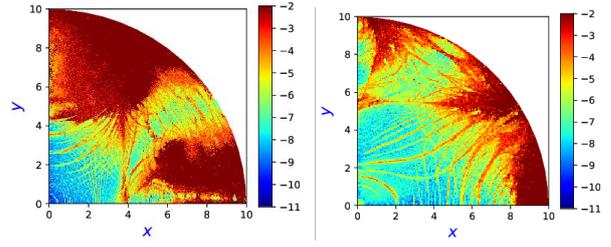

Figure 6: Amplitude space FMA plot for nominal tune (left) and a better tune (right) of (0.27, 0.26).

## INCREASING THE CROSSING ANGLE

The original crossing angle of 110 μrad together with β*=0.75 m results in a 12 σ separation at the parasitic interactions in the drift space before the first focusing quadrupole in the IR. To improve the DA, we increase the crossing angle to increase the initial separations from 12 σ to 20 σ. In addition, we also consider two other values of β*= (0.5, 1) m. Figure 7 shows the DA for different β* with increasing separation. For each β*, we find that the initial separations need to be 20 σ to reach the DA goal of 12 σ. Since the separation scales as 1/√β*, the crossing angles increase from 160 μrad at β*=1 m, to 184 μrad at β*=0.75 m and to 221 μrad at β*=0.5 m. For each β*, there is a 6 σ improvement in the DA from 12 σ separation to 20 σ separation. We find that the DA is independent on β* provided the scaled separation is constant. This may change when the nonlinear fields in the IR quadrupoles are included. In addition to beam dynamics constraints on the choices of β* and crossing angle such as luminosity, chromaticity correction, resonance excitation etc, the available physical aperture is very important. In the IR quadrupoles, the smallest physical aperture at constant initial separation of 20σ is 14σ for β*= 0.75m and drops to 9σ for β*= 0.5m. This could limit β* > 0.5m if the initial separation has to be 20σ.

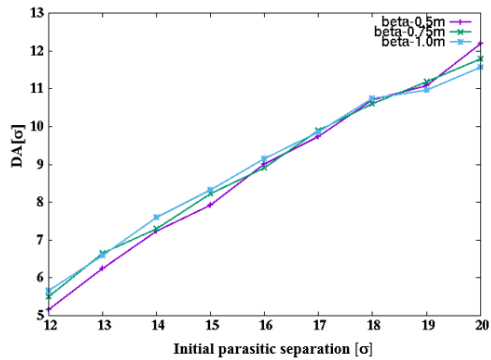

Figure 7: The DA for different β* with increasing 1st parasitic separation that is normalised by its horizontal beam size. The momentum deviation is 1 times rms momentum spread for all those cases.

## LONG-RANGE INTERACTION COMPENSATION

Here we examine the possibility of compensating the long-range interactions with current carrying wires [5]. In the lattice model, we have 4 wires distributed along the SPPC ring with one wire on each side in each IR. The quality of the compensation depends on the location of the wires. In [6] it was shown that ideally the phase advance between the long-range interactions (LRIs)), which are all at nearly the same phase, and the wire should be an integer multiple of π and the ratio of βx/βy should match that at the long-range interactions. However, this ratio varies from 1 over the first 12 of the parasitic interactions in the drift space to a high of 4.6 and a low of 0.2 over the remaining 29 interactions on one side of an IR. In the SPPC current lattice design, the phase advance between the LRIs to any place after the second separation dipole where the beam pipes are separate, is e.g. > ( 0.5π, 0.25 π) in the (x, y) planes respectively in the IR with horizontal crossing. Therefore, we have studied the compensation at the phase advance of nearly π in the crossing plane and at 3 different locations of the wire, the ratio on the right side of the IR with horizontal crossing has values of nearly βx/βy = (0.5, 1, 2). Simulations for the HL-LHC had found βx/βy=2 to be the most favourable [8]. Anti-symmetry of the optics about the IP implies that $(\beta x/\beta y)_{Right}$ = $(\beta y/\beta x)_{Left}$. When the phase advance is almost zero, the wire kick should have the opposite sign to those from the LRIs but when it is π, the wire kick should have the same sign [6, 7]. Figure 8 shows that the tune footprint for the baseline design is significantly smaller after we use the 4 wires. The wire is positioned at a normalized distance identical to the initial separation. Compensating all the interactions requires the current in each wire to be 118.1A.

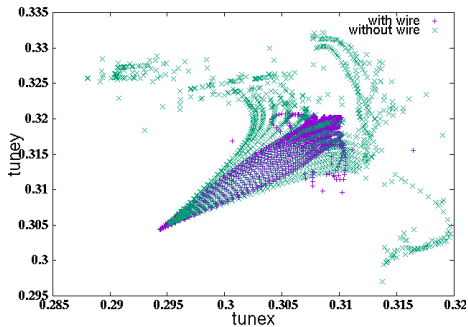

Figure 8: Tune footprint with wire compensation (violet points) and without compensation (green points)

The DA was calculated with the wires at the different locations. The phase advance changes little over these locations, so the phase advances are nearly π in the crossing plane in all cases. Figure 9 shows the DA as a function of the crossing angle before and after wire compensation. The wire's transverse position also changed with different crossing angle. This plot shows that the ratio βx/βy = 1 leads to the largest DA. This is because the beta ratio for all the LRIs in the drift sections is 1, therefore those LRIs will be better compensated with beta ratio 1 compared to beta ratio 0.5 and 2. We also observe that there's a 2σ increase in the DA with the wire compensation. This limitation of compensation effectiveness could be explained because the phase advance between LRIs and current wire is not exactly π and the beta ratio can only be partially matched. Considering the antisymmetry of the optics, when the beta ratio is either 2 or 0.5, the DA is about the same. FMA calculations do not reveal significant differences between the cases βx/βy=1 or 2. Long-term tracking also does not show large differences in the DA between the two cases, the largest is about 1σ when the initial separation is 15σ. This suggests that the phase advance may be most important in choosing optimum wire locations.

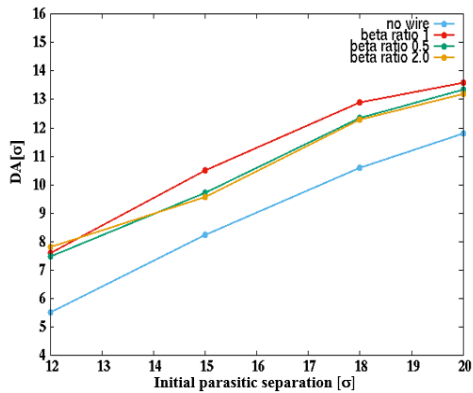

Figure 9: Dynamic aperture with increasing 1st parasitic separation that is normalised by its horizontal beam size for different beta ratios.

## CONCLUSIONS

Our simulations predict that the 164 long-range interactions are the main factor limiting particle stability and the DA in the SPPC baseline design is only 5.5σ. Tune scans showed that the beam-beam interactions strongly excite the 3rd, 4th, 5th, 9th and 10th order sum resonances, the latter two in the presence of the crossing angles. In addition, sychro-betatron coupling decreases the DA of off-momentum particles. The DA can be improved by ~1.5σ with different choices of tunes. To reach the DA goal of 12σ, increasing the crossing angle or increasing β* are the most useful options. Compensating the long-range interactions with current carrying wires is another option. We find that the optimal location for the wires is at a phase advance of π from the long-range interactions and where βx/βy = 1. At these locations the wires increase the DA by 2σ over a large range of crossing angles.